\documentclass[twoside]{article}

\usepackage{epsfig}
\usepackage{amssymb,amsmath,stmaryrd}

\catcode`\@=11
\long\def\@makefntext#1{
\protect\noindent \hbox to 3.2pt {\hskip-.9pt $^{{\eightrm\@thefnmark}}$\hfil}#1\hfill}  %CAN BE USED

\def\@makefnmark{\hbox to 0pt{$^{\@thefnmark}$\hss}}  %ORIGINAL

\def\ps@myheadings{\let\@mkboth\@gobbletwo
\def\@oddhead{\hbox{}
\rightmark\hfil\eightrm\thepage}
\def\@oddfoot{}\def\@evenhead{\eightrm\thepage\hfil
\leftmark\hbox{}}\def\@evenfoot{}
\def\sectionmark##1{}\def\subsectionmark##1{}}

\oddsidemargin=\evensidemargin
\newcounter{sectionc}\newcounter{subsectionc}\newcounter{subsubsectionc}
\renewcommand{\section}[1] {\vspace{12pt}\addtocounter{sectionc}{1}
\setcounter{subsectionc}{0}\setcounter{subsubsectionc}{0}\noindent
    {\tenbf\thesectionc. #1}\par\vspace{5pt}}
\renewcommand{\subsection}[1] {\vspace{12pt}\addtocounter{subsectionc}{1}
    \setcounter{subsubsectionc}{0}\noindent
        {\tenbf\thesectionc.\thesubsectionc. {\kern1pt\bfit #1}}
    \par\vspace{5pt}}
\renewcommand{\subsubsection}[1] {\vspace{12pt}\addtocounter{subsubsectionc}{1}
    \noindent{\tenrm\thesectionc.\thesubsectionc.\thesubsubsectionc.
    {\kern1pt \tenit #1}}\par\vspace{5pt}}
\newcommand{\nonumsection}[1] {\vspace{12pt}\noindent{\tenbf #1}
    \par\vspace{5pt}}

\newcounter{appendixc}
\newcounter{subappendixc}[appendixc]
\newcounter{subsubappendixc}[subappendixc]
\renewcommand{\thesubappendixc}{\Alph{appendixc}.\arabic{subappendixc}}
\renewcommand{\thesubsubappendixc}
    {\Alph{appendixc}.\arabic{subappendixc}.\arabic{subsubappendixc}}

\renewcommand{\appendix}[1] {\vspace{12pt}
        \refstepcounter{appendixc}
        \setcounter{figure}{0}
        \setcounter{table}{0}
        \setcounter{lemma}{0}
        \setcounter{theorem}{0}
        \setcounter{corollary}{0}
        \setcounter{definition}{0}
        \setcounter{equation}{0}
        \renewcommand{\thefigure}{\Alph{appendixc}.\arabic{figure}}
        \renewcommand{\thetable}{\Alph{appendixc}.\arabic{table}}
        \renewcommand{\theappendixc}{\Alph{appendixc}}
        \renewcommand{\thelemma}{\Alph{appendixc}.\arabic{lemma}}
        \renewcommand{\thetheorem}{\Alph{appendixc}.\arabic{theorem}}
        \renewcommand{\thedefinition}{\Alph{appendixc}.\arabic{definition}}
        \renewcommand{\thecorollary}{\Alph{appendixc}.\arabic{corollary}}
        \renewcommand{\theequation}{\Alph{appendixc}.\arabic{equation}}
        \noindent{\tenbf Appendix \theappendixc #1}\par\vspace{5pt}}
\newcommand{\subappendix}[1] {\vspace{12pt}
        \refstepcounter{subappendixc}
        \noindent{\bf Appendix \thesubappendixc. {\kern1pt \bfit #1}}
    \par\vspace{5pt}}
\newcommand{\subsubappendix}[1] {\vspace{12pt}
        \refstepcounter{subsubappendixc}
        \noindent{\rm Appendix \thesubsubappendixc. {\kern1pt \tenit #1}}
    \par\vspace{5pt}}

\newcommand{\textlineskip}{\baselineskip=13pt}
\newcommand{\smalllineskip}{\baselineskip=10pt}

\def\abstracts#1#2#3{{
    \centering{\begin{minipage}{4.5in}\baselineskip=10pt\footnotesize
    \parindent=0pt #1\par
    \parindent=15pt #2\par
    \parindent=15pt #3
    \end{minipage}}\par}}

\def\keywords#1{{
    \centering{\begin{minipage}{4.5in}\baselineskip=10pt\footnotesize
    {\footnotesize\it Keywords}\/: #1
    \end{minipage}}\par}}

\newcommand{\bibit}{\nineit}

\renewenvironment{thebibliography}[1]
    {\frenchspacing
     \ninerm\baselineskip=11pt
     \begin{list}{\arabic{enumi}.}
        {\usecounter{enumi}\setlength{\parsep}{0pt}
     \setlength{\leftmargin 12.7pt}{\rightmargin 0pt} %FOR 1--9 ITEMS
         \setlength{\itemsep}{0pt} \settowidth
    {\labelwidth}{#1.}\sloppy}}{\end{list}}

\newcounter{itemlistc}
\newcounter{romanlistc}
\newcounter{alphlistc}
\newcounter{arabiclistc}

\newcommand{\fcaption}[1]{
        \refstepcounter{figure}
        \setbox\@tempboxa = \hbox{\footnotesize Fig.~\thefigure. #1}
        \ifdim \wd\@tempboxa > 5in
           {\begin{center}
        \parbox{5in}{\footnotesize\smalllineskip Fig.~\thefigure. #1}
            \end{center}}
        \else
             {\begin{center}
             {\footnotesize Fig.~\thefigure. #1}
              \end{center}}
        \fi}

\newcommand{\tcaption}[1]{
        \refstepcounter{table}
        \setbox\@tempboxa = \hbox{\footnotesize Table~\thetable. #1}
        \ifdim \wd\@tempboxa > 5in
           {\begin{center}
        \parbox{5in}{\footnotesize\smalllineskip Table~\thetable. #1}
            \end{center}}
        \else
             {\begin{center}
             {\footnotesize Table~\thetable. #1}
              \end{center}}
        \fi}

\def\@citex[#1]#2{\if@filesw\immediate\write\@auxout
    {\string\citation{#2}}\fi
\def\@citea{}\@cite{\@for\@citeb:=#2\do
    {\@citea\def\@citea{,}\@ifundefined
    {b@\@citeb}{{\bf ?}\@warning
    {Citation `\@citeb' on page \thepage \space undefined}}
    {\csname b@\@citeb\endcsname}}}{#1}}

\newif\if@cghi
\def\cite{\@cghitrue\@ifnextchar [{\@tempswatrue
    \@citex}{\@tempswafalse\@citex[]}}
\def\citelow{\@cghifalse\@ifnextchar [{\@tempswatrue
    \@citex}{\@tempswafalse\@citex[]}}
\def\@cite#1#2{{$\null^{#1}$\if@tempswa\typeout
    {IJCGA warning: optional citation argument
    ignored: `#2'} \fi}}

\def\pmb#1{\setbox0=\hbox{#1}
    \kern-.025em\copy0\kern-\wd0
    \kern.05em\copy0\kern-\wd0
    \kern-.025em\raise.0433em\box0}

\def\fnt#1#2{\footnotetext{\kern-.3em           %FOOTNOTE TEXT
    {$^{\mbox{\sixrm\it #1}}$}{#2}}}

\def\fpage#1{\begingroup
\voffset=.3in
\thispagestyle{empty}\begin{table}[b]\centerline{\footnotesize #1}
    \end{table}\endgroup}

\def\runninghead#1#2{\pagestyle{myheadings}
\markboth{{\protect\footnotesize\it{\quad #1}}\hfill}
{\hfill{\protect\footnotesize\it{#2\quad}}}}
\font\tenrm=cmr10
\font\tenit=cmti10
\font\tenbf=cmbx10
\font\bfit=cmbxti10 at 10pt
\font\ninerm=cmr9
\font\nineit=cmti9

\font\eightrm=cmr8

\newtheorem{definition}{Definition}
\newtheorem{theorem}{Theorem}
\newtheorem{proposition}{Proposition}

\newtheorem{corollary}{Corollary}

\newcommand{\svee}{\operatornamewithlimits{\varovee}}
\newcommand{\swedge}{\operatornamewithlimits{\varowedge}}
\def\1{{\mathchoice {\rm 1\mskip-4mu l} {\rm 1\mskip-4mu l}
{\rm 1\mskip-4.5mu l} {\rm 1\mskip-5mu l}}}
\def \0{{\mathbb{O}}}

\def \half {\frac{1}{2}}
\newcommand{\sign}{\mathrm{sign}\,}

\newenvironment{proof}{\smallskip\noindent \bf Proof: \rm}
{\hspace*{\fill} $\square$ \newline} %\smallskip} %\blacksquare

\def\bsc{{\sc a\kern-6.4pt\sc a\kern-6.4pt\sc a}}   %LATEX LOGO
\def\bflatex{\bf L\kern-.30em\raise.3ex\hbox{\bsc}\kern-.14em
T\kern-.1667em\lower.7ex\hbox{E}\kern-.125em X}

\begin{document}
\runninghead{The quest for rings on bipolar scales}
{The quest for rings on bipolar scales}

\normalsize\textlineskip
\thispagestyle{empty}
\setcounter{page}{1}

%\copyrightheading{Vol. 0, No. 0 (2003) 000---000}

\vspace*{0.88truein} \fpage{1} \centerline{\bf The Quest for Rings on Bipolar Scales}
\vspace*{0.37truein}
\centerline{\footnotesize Michel GRABISCH$^1$, Bernard DE BAETS$^2$, and J\'anos FODOR$^3$}
\vspace*{0.055truein}
\centerline{\footnotesize\it $^1$Universit\'e Paris I --- Panth\'eon-Sorbonne}
\centerline{\footnotesize\it LIP6, 8 rue du Capitaine Scott, 75015 Paris, France}
\centerline{\footnotesize\it E-mail: \tt michel.grabisch@lip6.fr}
\vspace*{0.055truein}
\centerline{\footnotesize\it $^2$Department of Applied Mathematics, Biometrics and Process Control}
\centerline{\footnotesize\it Ghent University, Coupure links~653, B-9000 Gent, Belgium}
\centerline{\footnotesize\it E-mail: \tt Bernard.DeBaets@rug.ac.be}
\vspace*{0.055truein}
\centerline{\footnotesize\it $^3$Faculty of Veterinary Science}
\centerline{\footnotesize\it Szent Istv\'an University, Istv\'an u.~2., H-1078 Budapest, Hungary}
\centerline{\footnotesize\it E-mail: \tt jfodor@univet.hu}

\baselineskip=10pt
\vspace*{0.225truein}

%\publisher{March 2003}{.... 2003}

\vspace*{0.21truein} \abstracts{We consider the interval $]{-1},1[$ and intend to
endow it with an algebraic structure like a ring. The motivation lies in
decision making, where scales that are symmetric w.r.t.~$0$ are needed in order to
represent a kind of symmetry in the behaviour of the decision maker. A former proposal
due to Grabisch was based on maximum and minimum. In this paper, we propose to build
our structure on t-conorms and t-norms, and we relate this construction to uninorms. We show
that the only way to build a group is to use strict t-norms, and that there is no
way to build a ring. Lastly, we show that the main result of this paper is connected to the
theory of ordered Abelian groups.}{}{}

\vspace*{4pt}

\vspace*{18pt}\textlineskip

\section{Introduction}
So far, most of the studies of aggregation operators have been conducted on the unit
interval $[0,1]$, being representative for membership degrees for fuzzy sets,
uncertainty degrees for probability measures and other non-classical measures,
etc. From psychological studies, it is known that human beings handle other
kinds of scales. Three kinds of scales have been identified:
\begin{itemize}
\item {\em bounded unipolar scales}: typically $[0,1]$, suitable for bounded notions such as
membership degrees and uncertainty degrees;
\item {\em unipolar scales} (not necessarily bounded): typically $\mathbb{R}^+$,
suitable e.g.\ for priority degrees (one can always imagine something of higher priority);
\item {\em bipolar scales} (bounded or unbounded): typically $\mathbb{R}$, suitable for all paired
concepts of natural language, such as attraction/repulsion, good/bad, etc.
\end{itemize}
In decision making, it has been shown that the use of bipolar scales is of
particular interest, since they enable the representation of symmetry phenomena
in human behaviour, when one is faced with positive (gain, satisfaction, etc.)
or negative (loss, dissatisfaction, etc.) scores or utilities. This has led
for example to models in decision under uncertainty such as Cumulative Prospect
Theory\cite{tvka92}, based on the symmetric Choquet integral.

Recently, Grabisch has investigated symmetric algebraic structures
imitating the ring structure of real numbers, where the usual $+$ and $\times$ operations were
replaced by suitable extensions of maximum~$\vee$ and minimum~$\wedge$\cite{gra00,gra01d}.
The aim was to build an ordinal counterpart of numerical
bipolar scales, and of the symmetric Choquet integral, in order to make some first
steps towards an ordinal version of Cumulative Prospect Theory. He proposed new
operations, called symmetric maximum and minimum, and obtained a structure
close to a ring. In fact, associativity was problematic, but it was nevertheless
possible to define a symmetric Sugeno integral, counterpart of the symmetric
Choquet integral\cite{gra01e}.

Symmetric maximum and minimum can be seen as extensions of the usual $\vee$ and~$\wedge$
on $[{-1},1]$. Recall that $\vee$ (resp.~$\wedge$) is a t-conorm (resp. t-norm) on $[0,1]$.
The question that now
comes to mind is the following: Is it possible to extend any pair consisting of a t-conorm
and a t-norm to $[{-1},1]$ or $]{-1},1[$ so that the resulting structure is a ring? Or, at least, is it
possible to construct an Abelian group? A positive answer to this question would
permit readily to define symmetric fuzzy integrals defined with t-conorms and
t-norms\cite{musu91}. However, the results we present here indicate that the
quest for rings on bipolar scales seems to be a chimeric task.

\section{Background}
In this section, we introduce the concepts needed for our constructions. We refer the reader
to\cite{klmepa00} for a comprehensive treatment. It is also the source of our notations.

\vspace*{3pt}
\begin{definition}
A \emph{triangular norm} (t-norm for short) $T$ is a binary operation on $[0,1]$
such that for any $x,y,z\in [0,1]$ the following four axioms are satisfied:
\begin{itemize}
\itemindent=10pt
\item [\rm (\textbf{P1})] commutativity: $T(x,y)=T(y,x)$;
\item [\rm (\textbf{P2})] associativity: $T(x,T(y,z))= T(T(x,y),z))$;
\item [\rm (\textbf{P3})] monotonicity: $T(x,y)\leq T(x,z)$ whenever $y\leq z$;
\item [\rm (\textbf{P4})] neutral element: $T(1,x)=x$.
\end{itemize}
\end{definition}
Any t-norm satisfies $T(0,x)=0$. Typical t-norms are the minimum ($\wedge$),
the algebraic product~($\cdot$), and the {\L}ukasiewicz t-norm defined by
$T_{\mathbf{L}}(x,y):=(x+y-1)\vee 0$.

\vspace*{3pt}
\begin{definition}
A \emph{triangular conorm} (t-conorm for short) $S$ is a binary operation on $[0,1]$
that, for any $x,y,z\in [0,1]$, satisfies {\rm \textbf{P1}}, {\rm \textbf{P2}}, {\rm \textbf{P3}} and
\begin{itemize}
\itemindent=10pt
\item [\rm (\textbf{P5})] neutral element: $S(0,x) = x$.
\end{itemize}
\end{definition}

Any t-conorm satisfies $S(1,x)=1$. Typical t-conorms are the maximum $\vee$,
the probabilistic sum $S_\mathbf{P}(x,y):=x+y-xy$, and the {\L}ukasiewicz
t-conorm defined by $S_{\mathbf{L}}(x,y):=(x+y)\wedge 1$. T-norms and t-conorms
are dual operations in the sense that for any given t-norm $T$, the binary operation
$S_T$ defined by
$$
S_T(x,y) = 1-T(1-x,1-y)
$$
is a t-conorm (and similarly when starting from $S$). Hence, their properties are
also dual. The above examples are all dual pairs of t-norms and t-conorms.

A t-norm (or a t-conorm) is said to be \emph{strictly monotone} if $T(x,y)<T(x,z)$
whenever $x>0$ and $y<z$. A continuous t-norm (resp. t-conorm) is shown to be
\emph{Archimedean} if $T(x,x)<x$  (resp.~$S(x,x)>x$) for all $x\in\,]0,1[$.
A strictly monotone and continuous t-norm (resp. t-conorm) is called \emph{strict}. Strict t-norms
(resp. t-conorms) are Archimedean. Non-strict continuous Archimedean t-norms (resp. t-conorms)
are called \emph{nilpotent}.

Any continuous Archimedean t-conorm $S$ has an additive generator $s$, i.e. a strictly
increasing function $s:[0,1]\to [0,+\infty]$, with $s(0)=0$, such that, for
any $x,y\in [0,1]$:
\begin{equation}
\label{eq:tcon}
S(x,y) = s^{-1}[s(1)\wedge(s(x)+s(y))]\,.
\end{equation}
Similarly, any continuous Archimedean t-norm has an additive generator $t$ that is
strictly decreasing and satisfies $t(1)=0$. Strict t-conorms are characterized by
$s(1)=+\infty$, nilpotent t-conorms by a finite value of $s(1)$. Additive generators are
determined up to a positive multiplicative constant. If $t$ is an additive generator of a t-norm $T$,
then $s(x)=t(1-x)$ is an additive generator of its dual t-conorm~$S_T$.

\vspace*{4pt}
\begin{definition}\cite{yary96}
A \emph{uninorm} $U$ is a binary operation on $[0,1]$ that, for any $x,y,z\in [0,1]$, satisfies {\rm\textbf{P1}},
{\rm\textbf{P2}}, {\rm\textbf{P3}} and
\begin{itemize}
\itemindent=10pt
\item[\rm (\textbf{P6})] neutral element: there exists $e\in\,]0,1[$ such that $U(e,x)=x$.
\end{itemize}
\end{definition}

It follows that on $[0,e]^2$ a uninorm behaves like a t-norm, while on $[e,1]^2$
it behaves like a t-conorm. In the remaining parts, monotonicity implies that $U$ is comprised between
min and max. Associativity implies that $U(0,1)\in\{0,1\}$. Uninorms such that $U(0,1)=1$ are called disjunctive,
while the others are called conjunctive. By a simple rescaling, one can define from a given uninorm $U$ a
t-norm $T_U$ and a t-conorm $S_U$:
\begin{align}
T_U(x,y) & = \frac{1}{e}\,U(ex,ey) \label{eq:tu}\\
S_U(x,y) & = \frac{1}{1-e}\, \left[ U(e+(1-e)x,e+(1-e)y)-e \right]\,. \label{eq:su}
\end{align}
Conversely, from a given t-norm $T$ and t-conorm $S$, one can (partially) define a uninorm $U_{T,S}$ by:
\begin{equation}
U_{T,S}(x,y) = \left\{\begin{array}{ll}
        eT(\frac{x}{e},\frac{y}{e}) & \text{, if } (x,y)\in[0,e]^2\\
        e+(1-e)S\left(\frac{x-e}{1-e},\frac{y-e}{1-e}\right) & \text{, if }
        (x,y)\in [e,1]^2\,.
        \end{array} \right.
\end{equation}
The remaining parts can be filled in with minimum or maximum, leading to two extremal uninorms with the
given underlying operators. The problem of filling in the remaining parts is non-trivial and has been solved
recently for the case of a pair of continuous operators\cite{fobaca02}. Minimum and maximum, for instance,
can be extended in various ways to an idempotent uninorm\cite{bae99}.

\vspace*{4pt}
\begin{proposition}
\label{prop:uni}
The following statements are equivalent:
\begin{itemize}
\item[\rm (i)] $U$ is a uninorm with neutral element $e$, strictly monotone on $]0,1[^2$, and continuous
on $[0,1]^2\setminus\{(0,1),(1,0)\}$.
\item[\rm (ii)] There exists an additive generator $u$, i.e. a strictly increasing
$[0,1]\to [-\infty,\infty]$ mapping $u$ such that $u(e)=0$ and for any $x,y\in [0,1]$:
\begin{equation}
U(x,y) = u^{-1}(u(x)+u(y))\,,
\end{equation}
where by convention $\infty-\infty=-\infty$ if $U$ is conjunctive, and $+\infty$ if $U$ is disjunctive.
\end{itemize}
\end{proposition}
Uninorms characterized by the above proposition are called \emph{representable} uninorms.

Under the above assumption, it turns out that $T_U$ and $S_U$ are strict, and have
additive generators $t_u$ and $s_u$ defined by:
\begin{align}
t_u(x) & = -u(ex) \label{eq:tus}\\
s_u(x) & = u(e+(1-e)x)\,. \label{eq:sus}
\end{align}

\begin{definition}
Let $S$ be a t-conorm. The \emph{$S$-difference} is defined by:
\begin{equation}
\label{eq:diff}
x\ominus'_S y := \inf\{z \mid S(y,z)\geq x\}\,.
\end{equation}
\end{definition}
$S$-differences have been proposed by Weber\cite{web84}. From a logical point of view, $S$-differences
are extensions of the binary coimplication, the logical dual of implication (in the sense of de Morgan)\cite{bae97}.
$S$-differences are dual to residual implicators of t-norms.

If $S$ has an additive generator $s$, then it is easy to show that for any $x,y\in [0,1]$:
\begin{equation}
\label{eq:diff1}
x\ominus'_S y = s^{-1}(0\vee(s(x) - s(y)))\,.
\end{equation}
If $S=\max$, then this difference operator becomes:
\begin{equation}
\label{eq:diff3}
x\ominus'_\vee y = \left\{\begin{array}{ll}
            x & \text{, if } x>y\\
            0 & \text{, else}.
            \end{array} \right.
\end{equation}
In lattice theory, $x\ominus'_\vee y$ corresponds to the dual of the
pseudo-complement of $y$ relative to $x$ (see e.g.\cite{gratz98}).

\eject

\section{Symmetric pseudo-additions and pseudo-multiplications}\label{sec:spapm}
Following our motivation given in the introduction, we try to define in this
section what we mean by symmetric pseudo-addition and pseudo-multiplication on
$[{-1},1]$. Let us first consider the pseudo-addition.  Our aim is to endow
$[{-1},1]$ with a binary operation $\oplus$, extending a given t-conorm $S$, so that we
get a commutative group, with neutral element $0$.

A first basic fact is that we should consider the open interval $]{-1},1[$,
instead of the closed one if we want to get a group structure (and hence a
ring). Indeed, associativity implies that for any $a\in [0,1[$, we should have
$a\oplus(1\oplus -1) = (a\oplus 1)\oplus (-1)$. Since on $[0,1]$, $\oplus$
coincides with $S$, the above equality becomes $a\oplus(1\oplus -1) = 1\oplus
(-1)$, since 1 is an absorbing element. This equality cannot be satisfied for
all $a\in [0,1[$, unless $1\oplus(-1)$ is again an absorbing element; in
particular we cannot have $1\oplus(-1)=0$ as required by the group structure.
Hence, we exclude $-1$ and~$1$ to get a group, but we can still consider that $\oplus$
is defined on $[-1,1]^2$, imposing that $1\oplus(-1)$ is equal to $1$ or~$-1$. As
it will be seen in Section 5, the case of the closed interval
corresponds to \emph{extended} groups.

We propose the following definition.

\vspace*{4pt}
\begin{definition}
Given a t-conorm $S$, we define a \emph{symmetric pseudo-addition} $\oplus$
as a binary operation on $[-1,1]$:
\begin{itemize}
\itemindent=10pt
\item [\rm (\textbf{R1})] For $x,y\geq 0$: $x\oplus y = S(x,y)$.
\item [\rm (\textbf{R2})] For $x,y\leq 0$: $x\oplus y = -S(-x,-y)$.
\item [\rm (\textbf{R3})] For $x\in[0,1[$, $y\in\,]{-1},0]$: $x\oplus y = x\ominus_S (-y)$, where $\ominus_S$
is a symmetrized version of the $S$-difference:
$$
x\ominus_S y = \left\{\begin{array}{ll}
            \inf\{z \mid S(y,z)\geq x\} & \text{, if } x\geq y\\[.2cm]
            -\inf\{z \mid S(x,z)\geq y\} & \text{, if } x\leq y\\[.2cm]
            0 & \text{, if } x=y
            \end{array} \right.
$$
for $x,y\geq 0$. Moreover, $1\oplus(-1)=1$ or $-1$.
\item[\rm (\textbf{R4})] For $x\leq 0, y\geq 0$: just reverse $x$ and $y$.
\end{itemize}
\end{definition}
We now give justifications for our definition.
\begin{itemize}
\item On $[0,1]^2$, $\oplus$ has to be commutative, associative, have neutral
element $0$ and absorbing element $1$. Adding monotonicity (which may be
questionable), we have exactly the axioms of a t-conorm. This justifies
\textbf{R1}.
\item Since a group structure is desired, one should have a symmetric element
$-x$ for each element $x\in\, ]{-1},1[$, i.e. $x\oplus(-x)=0$. Using associativity
and commutativity, this leads to $0=(x\oplus (-x))\oplus(y\oplus(-y)) =
(x\oplus y) \oplus ((-x)\oplus (-y))$, which implies $(-x)\oplus (-y) =
-(x\oplus y)$, which in turn implies \textbf{R2}.
\item When $x\geq 0$ and $y\leq 0$, we define $x\oplus y:= x\ominus_S (-y)$. Although
this definition might seem to be somewhat arbitrary, the following
considerations justify this choice. For a strict t-conorm $S$ and $x\geq |y|$,
a comparison of formulas~(\ref{eq:diff1}) and~(\ref{eq:tcon}) clearly shows
that they complement each other. For the case of maximum, we refer to Section~4.3.
This justifies the first line of~\textbf{R3}.

Notice also that in the case of a representable uninorm $U$ with neutral
element $e\in\, ]0,1[$, and additive generator $u$, we always have $U(x, N(x))=e$
for $x\in\, ]0,1[$, where $N(x):=u^{-1}(-u(x))$ is a strong negation. After some
calculation and appropriate rescaling (i.e. the set $[e,1]\times [0,e]$ is
linearly mapped to $[0,1]\times[-1,0]$, whence $U$ becomes $\oplus$ and $N(y)$
becomes $-y$), one can come up with the formula
$$
x\oplus y = x\ominus'_S(-y)\,,
$$
for $(x,y)\in[0,1]\times [-1,0]$, $x\geq -y$. Compare this also with Section
4.1, especially Proposition~\ref{prop:u}.  The second line stems naturally
from the requirement $-(x\oplus y)= (-x)\oplus(-y)$, while the third line is due
to $x\oplus(-x)=0$.
\item \textbf{R4} originates from commutativity.
\end{itemize}

The case of pseudo-multiplication is less problematic. We want a commutative
and associative operation having 1 as neutral element and 0 as absorbing
element, and with the same rule of sign as the product, i.e.
$$
x\odot y = \sign(x\cdot y)\, T(|x|,|y|)\,,
$$
where the sign function is defined as usual by
$$
\sign(x) :=\left\{\begin{array}{ll}
            1 & \text{, if } x>0\\
            0 & \text{, if } x=0\\
            -1 & \text{, else.}
            \end{array}\right.
$$
This rule naturally follows from the distributivity of $\odot $ w.r.t. $\oplus$. Indeed,
$$
0=(x\oplus(-x))\odot y = (x\odot y) \oplus((-x)\odot y)\,,
$$
which entails $(-x)\odot y= - x\odot y$.

We propose the following definition.
\vspace*{4pt}
\begin{definition} Given a t-norm $T$, we define a \emph{symmetric pseudo-multiplication} $\odot$
as a binary operation on $[{-1},1]$:
\begin{itemize}
\itemindent=10pt
\item [\rm (\textbf{R5})] For $x,y\geq 0$: $x\odot y = T(x,y)$.
\item [\rm (\textbf{R6})] For other cases, $x\odot y = \sign(x\cdot y)\,T(|x|,|y|)$.
\end{itemize}
\end{definition}
Clearly, \textbf{R5} arises if we additionally require the monotonicity of $\odot$.
If the construction of $\oplus$ and $\odot$ can be done as described above,
and if in addition $\odot$ is distributive w.r.t. $\oplus$, then
$(]{-1},1[,\oplus,\odot)$ is a ring.

\section{Results}
We now examine the possibility of building symmetric pseudo-additions and multiplications leading
to a ring. We consider the following four cases, covering all continuous t-conorms:
\begin{enumerate}
\item $S$ is a strict t-conorm, with additive generator $s$;
\item $S$ is a nilpotent t-conorm, with additive generator $s$;
\item $S$ is the maximum operator;
\item $S$ is an ordinal sum of continuous Archimedean t-conorms (i.e. combination of the above).
\end{enumerate}

\subsection{Strict t-conorms}
\label{sec:ss}
Consider the symmetric pseudo-addition $\oplus$ defined from a strict t-conorm $S$ with
additive generator $s$. Let us rescale $\oplus$ to a binary operator $U$ on $[0,1]$ in the
following way:
\begin{equation}
\label{eq:trans}
x\oplus y = 2\,U\left(\half x+\half, \half y+\half\right)-1\,,
\end{equation}
and we put $z:=\half x+\half$, $t:=\half y+\half$ (coordinates in $[0,1]^2$).

Let us introduce $g:[-1,1]\to [-\infty,\infty]$ as follows:
\begin{equation}
\label{eq:syms}
g(x) := s(x) \text{ for positive }x, \quad g(x) := -s(-x) \text{ for negative
}x,
\end{equation}
i.e. $g$ is a symmetrization of $s$, and is clearly a strictly increasing
function. We introduce also another function
$u:[0,1]\to [-\infty,\infty]$ defined by $u(x) = g(2x-1)$ that is strictly
increasing and satisfies~$u(\half)=0$.

The following can be shown (originally established in\cite{klmepa96}).
\vspace*{4pt}
\begin{proposition}
\label{prop:u}
Let $S$ be a strict t-conorm with additive generator $s$, $\oplus$
the corresponding pseudo-addition, and $g$, $u$ defined as above. Then:
\begin{itemize}
\item [\rm (i)] $x\ominus y   = g^{-1}(g(x)-g(y))$ for any $x,y\in [0,1]$;
\item [\rm (ii)] $x\oplus y = g^{-1}(g(x)+g(y))$ for any $x,y\in [-1,1]$;
\item [\rm (iii)] $U(z,t) = u^{-1}(u(z)+u(t))$ for any $z,t\in [0,1]$.
\end{itemize}
\end{proposition}
\begin{proof}
(i) Follows immediately from the definition of $g$ and $\ominus$.

\noindent
(ii) Clearly $x\oplus y = s^{-1}(s(x)+s(y))=g^{-1}(g(x)+g(y))$ when $x,y\in
[0,1]$. When $x,y$ are negative, it holds due to \textbf{R2} that
\begin{align*}
x\oplus y & = - g^{-1}(g(|x|)+g(|y|))\\
    &  = g^{-1}(-(g(|x|)+g(|y|)))\\
    & = g^{-1}(g(x)-g(y))\,.
\end{align*}
When $x$ is positive and $y$ is negative, we write $x\oplus y = x\ominus (-y)$ and
use~(i). The result then follows from the symmetry of $g$.

\noindent
(iii) Using~(\ref{eq:trans}), the definition of $z$, $t$  and (ii), we get
\begin{align*}
2\,U(z,t)-1 & = g^{-1}(g(x)+g(y))\\
            & = g^{-1}(g(2z-1)+g(2t-1))\\
            & = g^{-1}(u(z)+u(t))\,.
\end{align*}
Now observe that for any $x$ and $y$, the equality $g^{-1}(y)=x$ is equivalent to
$y=g(x)=u(\half+\half x)$, which in turn is equivalent to $u^{-1}(y) =
\half+\half x=\half +\half g^{-1}(y)$. This shows that
$$
U(z,t) = \half g^{-1}(u(z)+u(t))+\half = u^{-1}(u(z)+u(t))\,.
$$
\end{proof}

\begin{corollary}
Under the assumptions of Proposition~\ref{prop:u}, $U$ is a uninorm that is
continuous (except in $(0,1)$ and $(1,0)$), is strictly increasing on $]0,1[^2$ and has
neutral element $\half$. Moreover, the t-norm $T_U$ induced is the dual of
$S$.
\end{corollary}

\begin{proof}
Since $u$ is strictly increasing and $u(\half)=0$, $U$ is a uninorm for which
statement~(ii) of Proposition~\ref{prop:uni} holds. Now, using (\ref{eq:tus}) and
(\ref{eq:sus}) with $e=\half$, we obtain $t(x)=-s(x-1)$, so that $T_U$ and $S$
are dual.
\end{proof}

Now we can state the main result.
\smallbreak
\begin{theorem}
\label{th:strict}
Let $S$ be a strict t-conorm with additive generator $s$ and
$\oplus$ the corresponding pseudo-addition. Then it holds that:
\begin{itemize}
\item[\rm (i)] $(]{-1},1[,\oplus)$ is an Abelian group.
\item[\rm (ii)] There exists no pseudo-multiplication such that
$(]{-1},1[,\oplus,\odot)$ is a ring.
\end{itemize}
\end{theorem}

\begin{proof}
(i) Since $\oplus$ is a uninorm with neutral element $\half$ transposed in
$[-1,1]^2$, commutativity and associativity hold, and the neutral element is
$0$. Moreover, symmetry holds since for any $x\in\, ]{-1},1[$ we can show that
\begin{align*}
x\oplus(-x) & = 2\,U\left( \half x+\half, -\half x+\half\right)-1\\
    & = 2u^{-1}(u(z)+u(1-z)) = 2u^{-1}(u(z)-u(z))-1 = 1
\end{align*}
using $u(z)=-u(1-z)$, for all $z\in[0,1]$.
\smallbreak
\noindent
(ii) By definition of the pseudo-multiplication, it is associative with
neutral element 1, so only distributivity w.r.t. $\oplus$ remains to be
verified.  If $T$ is the t-norm associated to $\oplus$, $T$ is distributive
w.r.t. $S$ if and only if $S=\max$\cite{klmepa00}. Since the maximum operator
is not a strict t-conorm, the construction cannot work.
\end{proof}

\subsection{Nilpotent t-conorms}
For a nilpotent t-conorm with additive generator $s$, it is easy to see that
statement~(i) in Proposition~\ref{prop:u} is still valid. However, the construction
does not lead to an associative operator, as it is easily seen in the following
example.  Consider the {\L}ukasiewicz t-conorm $S_{\mathbf{L}}(x,y)=(x+y)\wedge
1$ with additive generator $s_{\mathbf{L}}(x)=x$. Then $x\ominus y = x-y$. Let
$x=-0.3$, $y=0.6$, and $z=0.6$, then we obtain:
\begin{align*}
x\oplus(y\oplus z) & = -0.3\oplus(0.6\oplus 0.6) = -0.3\oplus 1 = 0.7\\
(x\oplus y)\oplus z & = (-0.3\oplus 0.6)\oplus 0.6 = 0.3\oplus 0.6 = 0.9\,.
\end{align*}
Hence, we can even not build a group in this case.

\subsection{The maximum operator}
\label{sec:max}
The case of maximum has already been studied by Grabisch in\cite{gra00}, in a
general setting where the underlying universe is a symmetric totally ordered
set, i.e. a structure $L=L^+\cup L^-$, where $L^+$ is a complete linear lattice
with top $\1$ and bottom $\0$, $L^-:=\{-x\mid x\in L^+\}$, and
$-x\leq -y$ if and only if $x\geq y$, with the convention $-\0=\0$. We are
interested in $L^+=[0,1]$, although subsequent results are valid in the general
case. The symmetric maximum and minimum, denoted $\svee$ and $\swedge$, are defined
as follows.
\begin{eqnarray}
a \svee b &:=& \left\{ \begin{array}{ll}
-(|a| \vee |b|) & \mbox{, if } b \neq -a \mbox{ and } |a| \vee |b|\in \{-a,-b\}\\
{\0} & \mbox{, if } b=-a \\
|a| \vee |b| & \mbox{, else} \end{array}\right.\label{eq:symmax}\\[.2cm]
a \swedge b &:=& \left\{ \begin{array}{ll}
-(|a| \wedge |b|) & \mbox{, if } \sign a \neq \sign b \\
|a| \wedge |b| & \mbox{, else.}
\end{array}
\right.\label{eq:symmin}
\end{eqnarray}
The structure $(]{-1},1[,\svee)$ fails to be a group since $\svee$ is not always associative.
For example, $$(0.5\svee 0.8)\svee(-0.8) = 0.8\svee (-0.8) = 0\,,$$
which differs from $$0.5\svee(0.8\svee(-0.8)) = 0.5\svee 0= 0.5\,.$$

A careful study in\cite{gra01d} has led to the following result.
Let us rephrase somewhat differently our requirements for symmetric maximum and
minimum:
\begin{itemize}
\itemindent=10pt
\item[(\textbf{C1})] $\svee$ and $\swedge$ coincide with $\vee$ and $\wedge$ on $L^+$, respectively;
\item[(\textbf{C2})] $\svee$ and $\swedge$ are associative and commutative on $L$;
\item[(\textbf{C3})] $-a$ is the symmetric element of $a$, i.e. $a\svee(-a) = \0$, for
$a\notin \{\1,-\1\}$;
\item[(\textbf{C4})] $-(a\svee b)=(-a)\svee(-b)$ and $-(a\swedge b)=(-a)\swedge
b$, for any $a,b\in L$.
\end{itemize}
Condition \textbf{C1} corresponds to \textbf{R1} and \textbf{R5}, while conditions
\textbf{C3} and \textbf{C4} correspond to \textbf{R2}, \textbf{R3} and
\textbf{R4}. Condition \textbf{C2} expresses group properties.
The following result shows that this task is impossible\cite{gra01d}.
\eject
\begin{proposition}
\label{prop:best}
We consider conditions {\rm(\textbf{C1})}, {\rm(\textbf{C3})}, {\rm (\textbf{C4})}, and
denote by {\rm(\textbf{C4+})} condition {\rm(\textbf{C4})} when $a$ and $b$ are restricted to
$L^+$. Then:
\begin{enumerate}
\item Conditions {\rm (\textbf{C1})} and {\rm (\textbf{C3})} imply that associativity
    cannot hold for $\svee$.
\item Under conditions~{\rm (\textbf{C1})} and~{\rm(\textbf{C4+})}, $\0$ is neutral for $\svee$. If we
    require in addition associativity, then $|a\svee(-a)|\geq |a|$. Further, if
    we require isotonicity of $\svee$, then $|a\svee(-a)|=|a|$.
\item Under conditions {\rm (\textbf{C1})}, {\rm (\textbf{C2})} and {\rm (\textbf{C3})},
no operation is associative on a larger domain than~$\svee$ defined by~(\ref{eq:symmax}).
\item $\varovee$ is associative for any expression involving
$a_1,\ldots,a_n$, $a_i\in L$, such  that $\bigvee_{i=1}^n a_i\neq
-\bigwedge_{i=1}^n a_i$.
\item The unique $\swedge$ satisfying {\rm (\textbf{C1})} and {\rm (\textbf{C4})} is
given by~(\ref{eq:symmin}), and is associative. $\swedge$ is distributive w.r.t. $\svee$
on $L^+$ and $L^-$ separately.
\end{enumerate}
\end{proposition}
The conflict between associativity (\textbf{C2}) and symmetry
(\textbf{C3}) is not surprising if we realize that for $S=\vee$, any element in
$]0,1]$ can be absorbing, if combined with smaller elements. Hence the basic
observation stated in the beginning of Section 3 applies.

Rescaling the symmetric maximum to a binary operator $U_{\max}$ on $[0,1]$
leads to:
\[
U_{\max}(x,y) :=\left\{\begin{array}{ll}
            \min(x,y) & \text{, if } x+y<1\\
            \frac12   & \text{, if } x+y=1\\
            \max(x,y) & \text{, if } x+y>1\,.
            \end{array}\right.
\]
As it lacks associativity, this operator is not a uninorm, but it is still very
much related to it. Indeed, the operator $U_{\max}$ coincides (except on the
antidiagonal $x+y=1$) with two typical idempotent uninorms (either taking min
or max on this antidiagonal)\cite{bae99}. However, it can also be obtained by
a limit procedure starting from a representable uninorm $U$
(by iteratively modifying its additive generator~$u$)\cite{meko98}.

\subsection{Ordinal sums}
Let us start with a simple case where $S$ is defined by a strict t-conorm $S_1$
in the upper corner $[a,1]^2$ for some $0<a<1$, and coincides with maximum elsewhere,
that is:
\begin{equation}
\label{eq:ords1}
S(x,y) = \left\{\begin{array}{ll}
        (1-a)\,S_1\left(\frac{x-a}{1-a},\frac{y-a}{1-a}\right)+a &
        \mbox{, if }x,y\in [a,1]\\ x\vee y & \text{, else}.
        \end{array} \right.
\end{equation}
Following our construction, we define $x\oplus y = S(x,y)$ on $[0,1]^2$ (this is \textbf{R1}).
Due to \textbf{R2}, \textbf{R3} and \textbf{R4}, it remains to define $x\ominus_S y$, for
$x,y\in [0,1]$, $x\geq y$. The following can be shown.

\eject
\begin{proposition}
Let $S$ be defined as in (\ref{eq:ords1}), and $s_1$ be an additive generator of $S_1$. Then
it holds that:
\begin{itemize}
\item[\rm (i)] For $(x,y)\in [a,1]^2$, $x\geq y$:
\begin{equation}
\label{eq:ords2}
x\ominus_S y =\left\{\begin{array}{ll}
        (1-a)s_1^{-1} \left(s_1\left(\frac{x-a}{1-a}\right)-
        s_1\left(\frac{y-a}{1-a}\right)\right) +a & \text{, if }x>y\\
        0 & \text{, else}\,;
            \end{array}\right.
\end{equation}
\item[\rm (ii)] For $(x,y)\in [0,1]^2\setminus [a,1]^2$, $x\geq y$:
\begin{equation}
x\ominus_S y = x\ominus'_\vee y\,.
\end{equation}
\end{itemize}
\end{proposition}

\begin{proof}
(i) By definition,
$$
x\ominus_S y = \inf\left\{z\geq a\mid S\left(\frac{y-a}{1-a},\frac{z-a}{1-a}\right)\geq
\frac{x-a}{1-a} \right\}\wedge \inf\{z<a \mid y\vee z\geq x\}\,.
$$
The second term does not exist unless $x=y$, and in this case it is 0. Since
$s_1$ is strictly increasing, the inequality in the first term is equivalent to
$$
\frac{z-a}{1-a}\geq s_1^{-1}\left(s_1\left(\frac{x-a}{1-a}\right)- s_1\left(\frac{y-a}{1-a}\right)\right)\,,
$$
and the result follows.\\
(ii) Observe that necessarily $y\leq a$. Then we have $$x\ominus_S y =
\inf\{z \mid y\vee z \geq x\}= x\ominus'_\vee y\,.$$
\end{proof}

Note that $\oplus$ is discontinuous on the line $x=y$, even if $x\geq a$.
Let us now express $\oplus$ using a generator function, where possible.
\vspace*{4pt}
\begin{proposition}
Let us define the generator function $g:[-1,-a[\,\cup\,\{0\}\,\cup\,]a,1]\to [-\infty,\infty]$ by:
$$
g(x) = \left\{\begin{array}{ll}
    s_1\left(\frac{x-a}{1-a}\right), & \text{if } x\in\, [a,1]\\
    -g(-x),  & \text{if } x\in [-1,-a[\\
    0, & \text{if } x=0
    \end{array}\right.
$$
with the convention $g^{-1}(0)=0$.
For any $x,y$ such that $(|x|,|y|)\in [a,1]^2$, we have:
$$
x\oplus y = g^{-1}(g(x)+g(y))\,.
$$
\end{proposition}
\begin{proof}
For $y> 0$, it holds that $g^{-1}(y) = x$ is equivalent to $y=g(x) =
s_1(\frac{x-a}{1-a})$, which in turn is equivalent to $s_1^{-1}(y) =
\frac{x-a}{1-a} = \frac{g^{-1}(y)-a}{1-a}$. Hence, we have $g^{-1}(y)=
s_1^{-1}(y)(1-a)+a$. Substituting in (\ref{eq:ords1}) and (\ref{eq:ords2}) and
considering symmetrization lead to the desired result for $x\neq- y$. In case
$x=-y$, we have $x\oplus y = g^{-1}(0) = 0$.
\end{proof}

Let us study the associativity of $\oplus$. Clearly, since an additive generator
exists, $\oplus$ is associative when the absolute value of the arguments belong to
$[a,1]$. Let us consider the problematic case $(b\oplus c)\ominus_S c$ when
$0<b<a<c<1$. We have $(b\oplus c)\ominus_S c = (b\vee c)\ominus_S c =
c\ominus_S c = 0$. However, $b\oplus (c\ominus_S c)=b$. Hence associativity
does not hold everywhere.

It is easily verified that putting a second strict t-conorm $S_2$ on $[0,a]^2$ in order
to cover the diagonal does not change the problem. Let us consider as above
$0<b<a<c<1$. Then $(b\oplus c)\ominus_S c=(b\vee c)\ominus_S c= 0$ again, and
associativity is not fulfilled. We conclude that ordinal sums cannot lead to associative operators,
although they result in ``more associative" operators than maximum does.

\section{On the relation with ordered Abelian groups}
\label{sec:oag}
The result on strict t-conorms could have been shown using results known in
the theory of ordered groups. We introduce the necessary definitions (see e.g.\cite{hahaji92,fuc63}).
\vspace*{4pt}
\begin{definition}
Let $(W,\leq)$ be a linearly ordered set, having top and bottom denoted
$\top$ and $\bot$, a particular nonextremal element $e$, and let us consider
an internal binary operation $\oplus$ on $W$, and a unary operation $\ominus$ on $W$ such that
$x\leq y$ if and only if $\ominus(x)\geq\ominus(y)$:
\begin{itemize}
\item[\rm (i)] $(W,\leq,\oplus,\ominus,e)$ is an \emph{ordered Abelian group} (OAG) if
 it satisfies for all \emph{nonextremal} elements $x,y,z$:
    \begin{itemize}
    \item[\rm (i1)] $x\oplus y = y\oplus x$;
    \item[\rm (i2)] $x\oplus(y\oplus z) = (x\oplus y)\oplus z$;
    \item[\rm (i3)] $x\oplus e=x$;
    \item[\rm (i4)] $x\oplus(\ominus(x))=e$;
    \item[\rm (i5)] $x\leq y$ implies $x\oplus z\leq y\oplus z$.
    \end{itemize}
\item[\rm (ii)] $(W,\leq,\oplus,\ominus,e)$ is an \emph{extended ordered Abelian group}
(OAG$^+$) if in addition
    \begin{enumerate}
    \item $\top\oplus x=\top$, $\bot\oplus x= \bot$ for all $x$,
$\ominus(\top)=\bot$, $\ominus(\bot)=\top$;
    \item if $x$ and $y$ are non-extremal, then $x\oplus y$ is non-extremal.
    \end{enumerate}
\end{itemize}
\end{definition}

Clearly, our concern is to find an OAG$^+$, with $W=[-1,1]$, $\top=1$, $\bot=-1$,
$\ominus=-$, and $\oplus$ corresponds to our operation $\oplus$.
\vspace*{4pt}
\begin{definition}
\begin{itemize}
\item[\rm (i)] An \emph{isomorphism} $\phi$ of an OAG (resp. OAG$^+$)
$\mathbf{W}=(W,\leq,\oplus,\ominus,e)$ onto an OAG (resp. OAG$^+$)
$\mathbf{W'}=(W',\leq',\oplus',\ominus',e')$ is a one-to-one mapping
from $W$ onto $W'$ preserving the structure, i.e. such that
\begin{itemize}
\item [\rm (i1)] $\phi(x\oplus y) = \phi(x)\oplus'\phi(y)$;
\item [\rm (i2)] $\phi(\ominus x) = \ominus'\phi(x)$;
\item [\rm (i3)] $\phi(e)=e'$;
\item [\rm (i4)] $x\leq y$ if and only if $\phi(x)\leq' \phi(y)$.
\end{itemize}
\item[\rm (ii)] $\mathbf{W}$ is a \emph{substructure} of $\mathbf{W'}$ if
$W\subset W'$ and the structure of $\mathbf{W}$ is the restriction of the structure of
$\mathbf{W'}$ to $W$, i.e. $x\oplus y=x\oplus' y$, $\ominus x=\ominus'x$,
$e=e'$, and $x\leq y$ if and only if $x\leq' y$, for all $x,y\in W$.
\item[\rm (iii)] An \emph{isomorphic embedding} of $\mathbf{W}$ into $\mathbf{W'}$
is an isomorphism of $\mathbf{W}$ onto a substructure of $\mathbf{W'}$.
\end{itemize}
\end{definition}
\smallbreak
Observe that in (i) above, in the case of OAG$^+$, we have in addition
$\phi(\top) = \top'$ (due to (i1) with $x=\top$), and $\phi(\bot) = \bot'$~(due to
(i2)).
\vspace*{4pt}

\begin{definition}
\begin{itemize}
\item[\rm (i)] An OAG $\mathbf{W}$ is \emph{dense} if there is no least positive element,
i.e. an element $x\in W$ such that $x>e$, and there is no $y\in W$ such that
$e<y<x$.
\item[\rm (ii)] An OAG $\mathbf{W}$ is \emph{completely ordered} if each non-empty
bounded $X\subset W$ has a least upper bound.
\end{itemize}
\end{definition}
Obviously, $(]{-1},1[,\leq,\oplus,-,0)$ is dense and completely ordered, the
same holds for the closed interval.
\vspace*{4pt}
\begin{theorem}
If $\mathbf{W}$ is a completely ordered and dense OAG, then it is isomorphic to
$(\mathbb{R},\leq,+,-,0)$.
\end{theorem}
The same result holds if $\mathbf{W}$ is an OAG$^+$ and if $\mathbb{R}$ is
replaced by $\overline{\mathbb{R}}:=\mathbb{R}\cup\{-\infty,\infty\}$.
This shows that $\oplus$ has necessarily the following form:
\begin{equation}
x\oplus y = \phi^{-1}[\phi(x)+\phi(y)]\,,
\end{equation}
where $\phi:[-1,1]\to \overline{\mathbb{R}}$ is one-to-one,
odd, increasing, and satisfies $\phi(0)=0$. Clearly, $\phi$ corresponds to the
function $g$ of Proposition~\ref{prop:u}.

\section{Concluding remarks}
In this paper we have studied whether it is possible to introduce a ring
structure on $]{-1},1[$, using a t-conorm and a t-norm. The results show that the
answer is negative. In the particular case of the maximum and minimum, the
symmetric maximum and minimum\cite{gra00} are the best possible choices: they
provide a structure that is not a ring, but is the closest possible to a ring
in a well-defined sense.

One may be interested in finding a ring structure on finite ordinal
scales. However, it is well known that there exist no strict t-conorms on finite
ordinal scales\cite{mato93}. Hence, there is no chance to obtain an Abelian
group. Nilpotent t-conorms lead to disadvantageous results as we have shown
above. Hence, one can conclude that the best choice seems to be the symmetric
maximum.

Finally, we remark that this structure together with the symmetric minimum is the
closest possible to a ring, due to the fact that the only chance to get
distributivity is with $S = \max$.

\eject
\nonumsection{Acknowledgements}
\noindent
This work is supported in part by the Bilateral Scientific and Technological
Cooperation Flanders--Hungary BIL00/51 (B-08/2000), by FKFP 0051/2000, and by
OTKA 046762.

\nonumsection{References}
\vspace*{-10pt}
\noindent

\bibliographystyle{plain}

\end{document}